\documentclass[letter]{aa}

\usepackage{graphicx}
\usepackage{txfonts}
\usepackage{mathtools}
\usepackage{bm}
\usepackage{natbib}
\bibpunct{(}{)}{;}{a}{}{,} 
\usepackage{hyperref}
\usepackage{float}

\newcommand{\e}{\operatorname{e}}

\begin{document}

   \title{The slope of the power spectrum of the density field in isothermal supersonic compressible turbulence}
   \titlerunning{The slope of the PS of the turbulent density field}

   \author{Pierre Dumond \inst{1}, Jérémy Fensch \inst{1}, Gilles Chabrier\inst{1,2} \and Noé Brucy\inst{1} }

   \institute{ENS de Lyon, CRAL UMR5574, Universite Claude Bernard Lyon 1, CNRS, Lyon 69007, France
         \and
             School of Physics, University of Exeter, Exeter, EX4 4QL, UK
             }

   \date{Received XXX; accepted XXX}

  \abstract{ 
    The power spectrum (PS) of the density field in supersonic turbulence is a fundamental quantity that characterizes the statistical properties of the structures formed in compressible flows. It is also widely used to estimate the Mach number in the interstellar medium from simulation-derived relations. In this paper, we provide a first quantitative explanation for the evolution of the slope of the PS of the density field with the Mach number in homogeneous isotropic isothermal turbulence using a time-invariant quantity derived by Chandrasekhar (1951). For simulated turbulent flows, the model reproduces the measured slopes for different widths of the inertial range and density variances very well. Our model also provides a comprehensive interpretation of the characteristic slopes of the PS of the density field measured in the interstellar medium. Based on these results, we stress that the Mach number cannot be reliably deduced from the slope of the PS of the density field. In closing, we discuss a resolution criterion that must be fulfilled to correctly simulate a turbulent flow with a given density PS slope.
   }

   \keywords{Methods: analytical --- Turbulence
               }

   \maketitle
%
\nolinenumbers

\section{Introduction}

The statistical properties of supersonic compressible turbulent flows are not well predicted from first principles because the typical closure equations cannot be applied to such highly non-linear flows \citep{Bertoglio_TwopointClosuresWeakly2001, Zhou_TurbulenceTheoriesStatistical2021}. Therefore, their knowledge is mainly based on phenomenological models. For example, the 3D power spectrum (PS)\footnote{In this study, we always consider the PS in three dimensions.} of the velocity field is known to be proportional to \(k^{-4}\) in the inertial range for supersonic compressible turbulent flows \citep{Burgers_MathematicalModelIllustrating1948}. It is interpreted as the consequence of the discontinuities created by supersonic shocks. 
However, the PS of the density field, though it is usually observed with a slope close to \(k^{-3}\) both in simulations of supersonic turbulence \citep{Kim_DensityPowerSpectrum2005, Konstandin_MachNumberStudy2016} and in the supersonic interstellar medium \citep{Pingel_MultiphaseTurbulenceDensity2018}, has not been explained so far. 
Simulations \citep{Kim_DensityPowerSpectrum2005, Konstandin_MachNumberStudy2016, Pan_ExactRelationDensity2022} usually show that its power-law exponent flattens as the Mach number increases. A similar evolution has also been observed in laboratory experiments \citep{White_SupersonicPlasmaTurbulence2019}.
\citep{Squire_DistributionDensitySupersonic2017} has suggested a phenomenological model of shocks that takes into account the intermittency of the turbulent flow and that predicts the power-law exponent in the high Mach number limit. However, their model strongly depends on ill-constrained parameters. \cite{Rastegar_ExactTwopointCorrelation1998} has also developed a model that neglects the pressure gradient, which could be a valid approximation in the high Mach number limit, and found a slope close to -3.
None of these models predict the evolution of the power-law exponent with the Mach number.

In the astrophysical context, the PS of the density field can serve as a probe of the evolutionary stage of molecular clouds \citep{Federrath_STARFORMATIONEFFICIENCY2013, Burkhart_OBSERVATIONALDIAGNOSTICSSELFGRAVITATING2015}. A shallow slope suggests that a significant amount of gas has collapsed into stars, while a steeper slope suggests that the cloud is still at the beginning of its evolution and that star formation has not started. 
In this paper, based on a time invariant quantity that holds in statistically homogeneous compressible turbulence (\citealt{Chandrasekhar_FluctuationsDensityIsotropic1951,Jaupart_EvolutionDensityPDF2020, Dumond_MassInvariantCompressible2025}; D25 hereafter), we derive a model that relates the slope of the density PS to the variance of the density field in the case of isothermal turbulence.

\section{Model of the power spectrum of the density field}

\subsection{The mass invariant} 
Assuming ergodicity, and thus statistical homogeneity of the turbulent flow, \cite{Chandrasekhar_FluctuationsDensityIsotropic1951} and  \cite{Jaupart_GeneralizedTransportEquation2021a} derived a time invariant based on the continuity equation. This invariant holds if and only if the cross-correlation function, \(C_{\rho, \rho\bm{v}}(|\bm{q}|)\), between the density, \(\rho\), and the momentum, \(\rho\bm{v}\), decays faster than \(1/|\bm{q}|^2\) at infinity, where \(\bm{q}\) denotes the spatial distance.
This invariant is defined as
\begin{equation}
    \label{eq_M_inv_def}
    M_{\rm inv} = \mathbb{E}(\rho)(t)\text{Var}\left(\frac{\rho}{\mathbb{E}(\rho)}\right)_t(l_{\rm c}^\rho)_t^3 = \text{const},
\end{equation}
where \(l_{\rm c}^\rho\) is the correlation length of the density field \citep{Jaupart_GeneralizedTransportEquation2021a,Jaupart_StatisticalPropertiesCorrelation2022} and \(\mathbb{E}(\rho)\) and \({\rm Var}(\rho)\) are the expectation value and the variance of the density field, respectively.

This time invariance has been verified numerically in the case of decaying turbulence for various Mach numbers and equations of state in D25 (see also Appendix \ref{App_numerical_setup}).
Moreover, D25 also showed that this time-invariance property implies that \(M_{\rm inv}\) is Mach invariant in the supersonic regime (see discussion in D25, Sect. V.a).
This Mach invariance of \(M_{\rm inv}\) has been confirmed numerically for a log-density variance of \(\sigma_s^2\lesssim 5\), corresponding to a Mach number of \(\mathcal{M}\lesssim 10\).

The invariant does not depend on the dissipation scale either because the latter will have no impact on the cross-correlation between the density and momentum at large scales. The invariant is thus not modified by variations of the dissipation scale.

To determine the value of the invariant, we relied on the model of the log-density field presented in D25 and recalled in Appendix~\ref{App_logdensity_PS}. Assuming that in the transonic limit (\(\mathcal{M}\sim 2\)), the non-Gaussian features of the log-density field are small, we get 
\[ M_{\rm inv} = \alpha M_{\rm inj},\] 
with \(\alpha= 2 \times 10^{-2}\) and \(M_{\rm inj} = \mathbb{E}(\rho)L_{\rm inj}^3\), where \(L_{\rm inj}\) is the scale of maximum energy injection. This value is obtained without any fitting parameter and is consistent with the value of \(\alpha\) measured within the 1\(\sigma\) uncertainties in the numerical simulations of D25. 
As discussed in D25, the value of \(\alpha\) predicted by the model retains some uncertainty due to the lack of constraints on the Mach number to be used to determine \(M_{\rm inv}\) from the log-density statistics. Although this will affect the prediction of \(\alpha\) by at most 20\% and the prediction of the slope of the density PS by the model by at most 5\%, it will not affect the qualitative evolution of the slope with the Mach number.

\subsection{Functional form of the power spectrum of the density field} 
\begin{figure}
    \centering
    \includegraphics[width=\columnwidth]{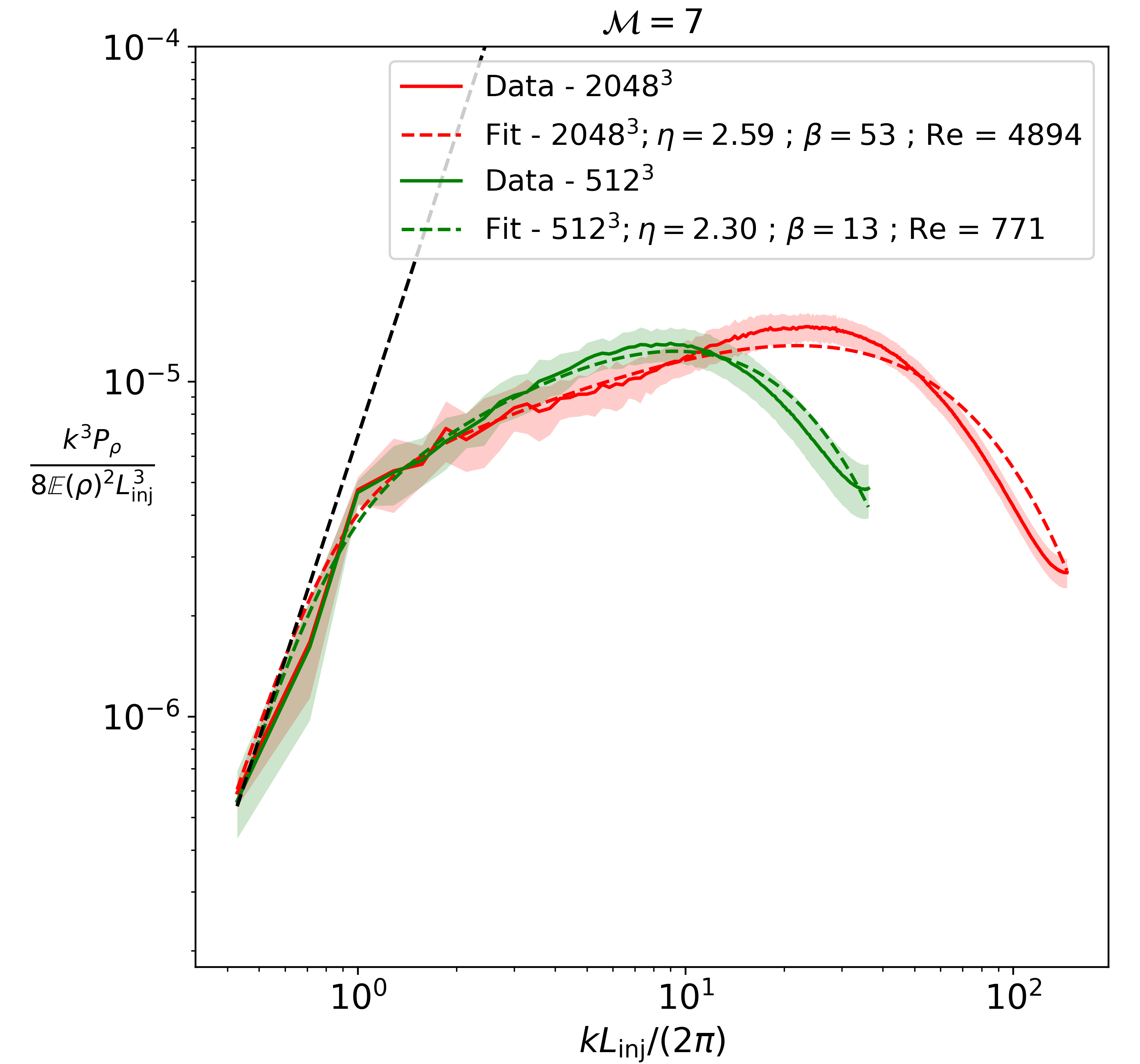}
    \caption{Measured PS of the density field at \(\mathcal{M}=7\) and its fit with Eq. \ref{eq_fit_PS_rho} with \(p=2\) for two different resolutions. The PS has been normalized by \(8\mathbb{E}(\rho)^2 L_{\rm inj}^3\) such that its asymptotic value at \(k\to 0\) corresponds to \(k^3\alpha\). The dashed black  line corresponds to \(k^3\alpha\), with \(\alpha=2\times 10^{-2}\), which is the value predicted by our model (see also D25). The shaded area corresponds to the 1\(\sigma\) uncertainties estimated from the time variations of the density power spectra. The inset of the left panel shows the projected density field for the 512\(^3\) resolution.}
    \label{fig_PS_fit}
\end{figure}

By analogy with the Richardson cascade \citep{Richardson_WeatherPredictionNumerical1922}, we parametrized the PS of the density field with the following functional form:
\begin{align}
    \label{eq_fit_PS_rho}
    P_\rho^{\rm 3D}(k) = \text{Var}\left(\frac{\rho}{\mathbb{E}(\rho)}\right)\frac{A}{\left(1+\left(\frac{kL_{\rm inj}^{\rm max}}{2\pi}\right)^{2p}\right)^\frac{\eta}{2p}} \e^{-k \frac{L_{\rm diss}}{2\pi}}.
\end{align}
Since this functional form is not derived from first principles, we introduced an additional parameter, $p$, which controls the sharpness of the transition between the injection range and the inertial range.
Furthermore, this formulation involves $L_{\rm inj}^{\rm max}$, the largest scale at which forcing is applied to the medium.
This functional form models the three ranges that characterize the PS of the density field. At scales larger than the injection scale (\(2\pi/k>L_{\rm inj}^{\rm max}\)), the density field is completely uncorrelated, and the PS is expected to be flat.
Between the injection scale and the dissipation scale, the inertial range is characterized by a power law with a slope \(-\eta\) that we aim to determine. For scales smaller than \(L_{\rm diss}\), the PS decays exponentially in the dissipation range in analogy with the functional form of the dissipation range of the velocity power spectrum \citep{Burgers_MathematicalModelIllustrating1948,Frisch_TurbulenceLegacyKolmogorov1995}. 
We also introduced the normalization coefficient \(A\), which ensures that the integral of the PS is equal to the variance of the density field (see Eq. \ref{eq_Parseval}). 
In Fig.~\ref{fig_PS_fit}, we present the density power spectrum measured in our numerical simulations together with its fit using Eq.~\ref{eq_fit_PS_rho} with $p=2$. {The power spectrum was computed from the square of the Fourier transform (FT) of the fluctuations of the density field, \(\rho-\mathbb{E}(\rho)\). From the Wiener-Khintchine theorem, this is equivalent to the FT of the autocovariance function for \(k\neq 0\).}
We find that this functional form provides a very good description of the measured power spectrum. In Appendix~\ref{App_dependance_func_PS_rho}, we show that our results depend only weakly on the chosen value of $p$ for $p \geq 2$. For smaller values of $p$, the transition between the injection and inertial ranges becomes less rapid and does not accurately reproduce the power spectrum obtained from the numerical simulations.

\subsection{Prediction of the slope of the density power spectrum} 
Due to the non-Gaussian features of the log-density field increasing with Mach number (\citealt{Hopkins_ModelNonlognormalDensity2013a,Squire_DistributionDensitySupersonic2017}, D25) and  the flattening of the log-density PS at high Mach numbers \citep{Brucy_InefficientStarFormation2024a} making Eq. \ref{eq_PS_logdensity_diss} increasingly inaccurate, the model of the statistics of the log-density field (Appendix~\ref{App_logdensity_PS}) cannot be used to compute the PS of the density field directly.
To bypass this limitation, we used the aforementioned invariant to predict the slope of the density PS in the supersonic regime. 
As shown in Appendix \ref{App_analytical_details}, we can relate the density variance to the slope \(\eta\):
\begin{equation}
    \label{eq_slope_variance}
    \text{Var}\left(\frac{\rho}{\mathbb{E}(\rho)}\right) = 32\pi \alpha \int_{0}^{\infty}\frac{u^2 \e^{-u\beta^{-1}}}{(1+(\phi u)^2)^\frac{\eta}{2}}{\rm d}u,
\end{equation}
where we have introduced \(\beta=L_{\rm inj}/L_{\rm diss}\), which quantifies the width of the inertial range and \(\phi = L_{\rm inj}^{\rm max}/L_{\rm inj}\).
Because \(\alpha\) does not depend on the density variance nor on the dissipation scale and \(\phi\) is imposed by the forcing properties of the system, the model shows that the slope \(\eta\) depends on both the density variance, \(\text{Var}\left(\frac{\rho}{\mathbb{E}(\rho)}\right)\), and the inertial width, \(\beta\).

\subsection{Comparison with numerical simulations} 

The data used to compare our predictions to numerical experiments are taken from the simulations presented in D25 and detailed in Appendix \ref{App_numerical_setup}. These are uniform grid simulations run with the hydrodynamical code RAMSES \citep{Teyssier_CosmologicalHydrodynamicsAdaptive2002}. The numerical method is based on a second-order Godunov solver scheme, and the turbulence was forced using an Ornstein-Uhlenbeck process on the acceleration in Fourier space.
The originality of this set of simulations is that turbulence is forced at scales significantly smaller than the box scale (between \(L_{\rm box}/6\) and \(L_{\rm box}/8\)), which ensures that the statistical estimates of the correlation length, of the variance, and of \(M_{\rm inv}\) are robust as shown in D25. 

To determine the slope of the PS of the density field, we fit the measured PS of the density field with the functional form of Eq. \ref{eq_fit_PS_rho}, with \(\eta\) being the only free parameter. Because the dissipation range has been found to be universal in grid simulations, where the dissipation is induced by the grid \citep{Federrath_UniversalitySupersonicTurbulence2013}, we used the best-fitting value of \(L_{\rm diss}\) normalized to the spatial resolution \(\Delta x\): \(L_{\rm diss} = 5.5 \Delta x\). This value is valid for the numerical solver used in the present study.

In Fig. \ref{fig_PS_fit}, we plot our measured PS (solid) and the corresponding fit (dashed) for two different resolutions.
This independence of \(\alpha\) from the spatial resolution is illustrated by the fact that \(\lim_{k\to 0}P_\rho(k)\) does not vary within the 1\(\sigma\) uncertainties illustrated by the shaded area.

\begin{figure}
    \centering
    \includegraphics[width=\columnwidth]{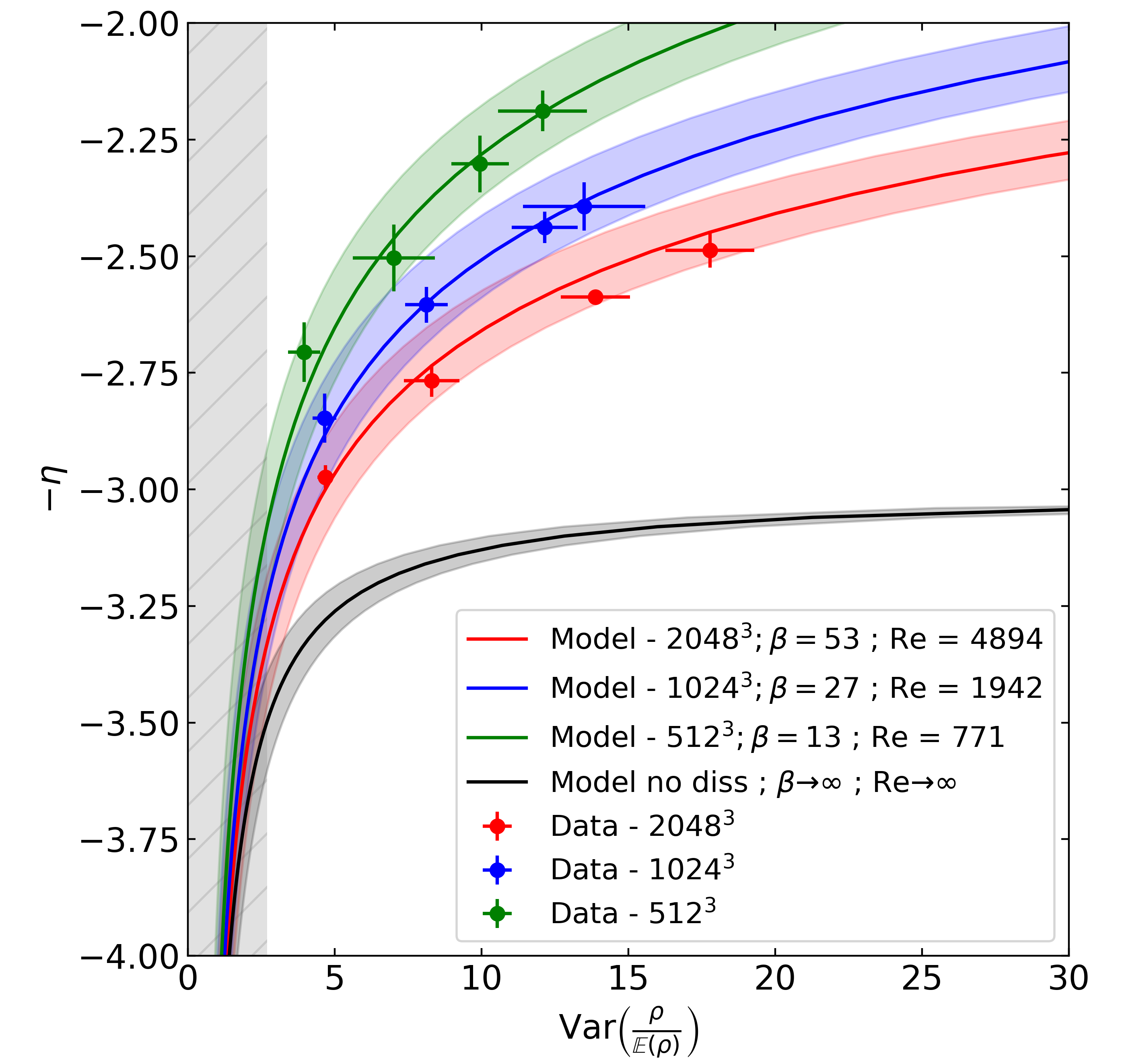}
    \caption{Predicted (solid lines) and measured (dots) slope \(-\eta\) in the inertial range of the PS of the density field with the variance at different resolutions, taking \(p=2\). The black curve corresponds to the predicted evolution of the slope in the absence of dissipation (\(L_{\rm diss}=0\)). The gray shaded area corresponds to the region where the model does not predict the evolution of the slope because the invariance of \(M_{\rm inv}\) is not verified for such a small variance. The error bars correspond to the $\pm 1\sigma$ uncertainty estimated from the time variations of the measured quantities. The shaded areas around the solid curves correspond to different predictions of the model considering an uncertainty of 20\% on the determination of the invariant around its predicted value \(\alpha=2\times 10^{-2}\). }
    \label{fig_slope_model}
\end{figure}

In Fig. \ref{fig_slope_model}, we plot the evolution of the slope of the PS of the density field predicted by the model (Eq. \ref{eq_slope_variance}) with the density variance for several dissipation lengths. 
We compare these predictions with the slope of the PS of the density field measured in the simulations of D25.
The model predicts the evolution of the slope of the PS with the density variance very well.

In the limit of zero viscosity (\(\beta\to\infty\)), the model predicts that the slope of the PS of the density field tends to -3 in the high-variance limit. This result is consistent with the theoretical result of \cite{Rastegar_ExactTwopointCorrelation1998} that predicts a slope of -2.95 in the pressureless limit (the sound speed tends to 0), which is equivalent to the high Mach number (or high variance) limit. In Appendix \ref{App_no_comp} we explain why we do not compare our prediction to other numerical studies.

This model highlights the fact that the variation of the slope of the density PS with the Mach number observed in numerical simulations \citep{Kim_DensityPowerSpectrum2005,Konstandin_MachNumberStudy2016} is a direct consequence of the dissipation scale imposed by the grid. In incompressible turbulence, for instance, the dissipation scale \(L_{\rm diss}\) depends on the velocity dispersion \(\sigma_v\). In realistic supersonic compressible turbulence, the dependence of the dissipation scale on the Mach number is unconstrained but has no reason to be constant with respect to the Mach number. 

{In Appendix \ref{App_explicit_visc} we show that adding an explicit viscosity, which reduces the inertial range without changing the resolution, modifies the slope of the density power spectrum similarly to grid viscosity. This confirms that the slope depends on viscosity. 
Furthermore, if the slope of the power spectrum was independent of viscosity, as is commonly assumed in turbulence, the slope should be steeper than -3. However, plasma experiments report density power spectra with slopes shallower than \(-3\) \citep{White_SupersonicPlasmaTurbulence2019}. This supports the conclusion that the slope of the density power spectrum depends on viscosity.
}

\section{Application to astrophysical systems: Turbulent molecular clouds} 
Here, we compare the slope of the observed PS in molecular clouds and our prediction.
Such a comparison, however, is difficult because neither the variance of the volume density field nor the ratio of the dissipation scale over the injection scale are well constrained.
It is possible to get an estimate of the variance from the Mach number using the first-order relation \(\text{Var}(\rho/\mathbb{E}(\rho)) = b^2 \mathcal{M}^2\) (see, e.g., \citealt{Konstandin_MachNumberStudy2016}). Taking \(b=0.5\), which corresponds to an equipartition forcing between solenoidal and compressive modes \citep{Federrath_ComparingStatisticsInterstellar2010}, and a typical Mach number ranging between 4 and 20 for star-forming clouds \citep{Larson_TurbulenceStarFormation1981, Chabrier_VariationsStellarInitial2014} leads to a variance ranging between 4 and 100 \citep{Hennebelle_TurbulentMolecularClouds2012}.

Moreover, the processes that dissipate turbulence in molecular clouds are not well identified. Although it has been suggested that the main dissipation process of turbulence at small scales is ambipolar diffusion (see, e.g., \citealt{Momferratos_TurbulentEnergyDissipation2014}), this view has been challenged recently \citep{ Pineda_ProbingPhysicsStarFormation2024a}. 
The fact that the power law of the density PS is observed over more than three decades in spatial scales means that the inertial width is \(\beta>10^{3}\) for such systems.

\begin{figure}
    \centering
    \includegraphics[width=\columnwidth]{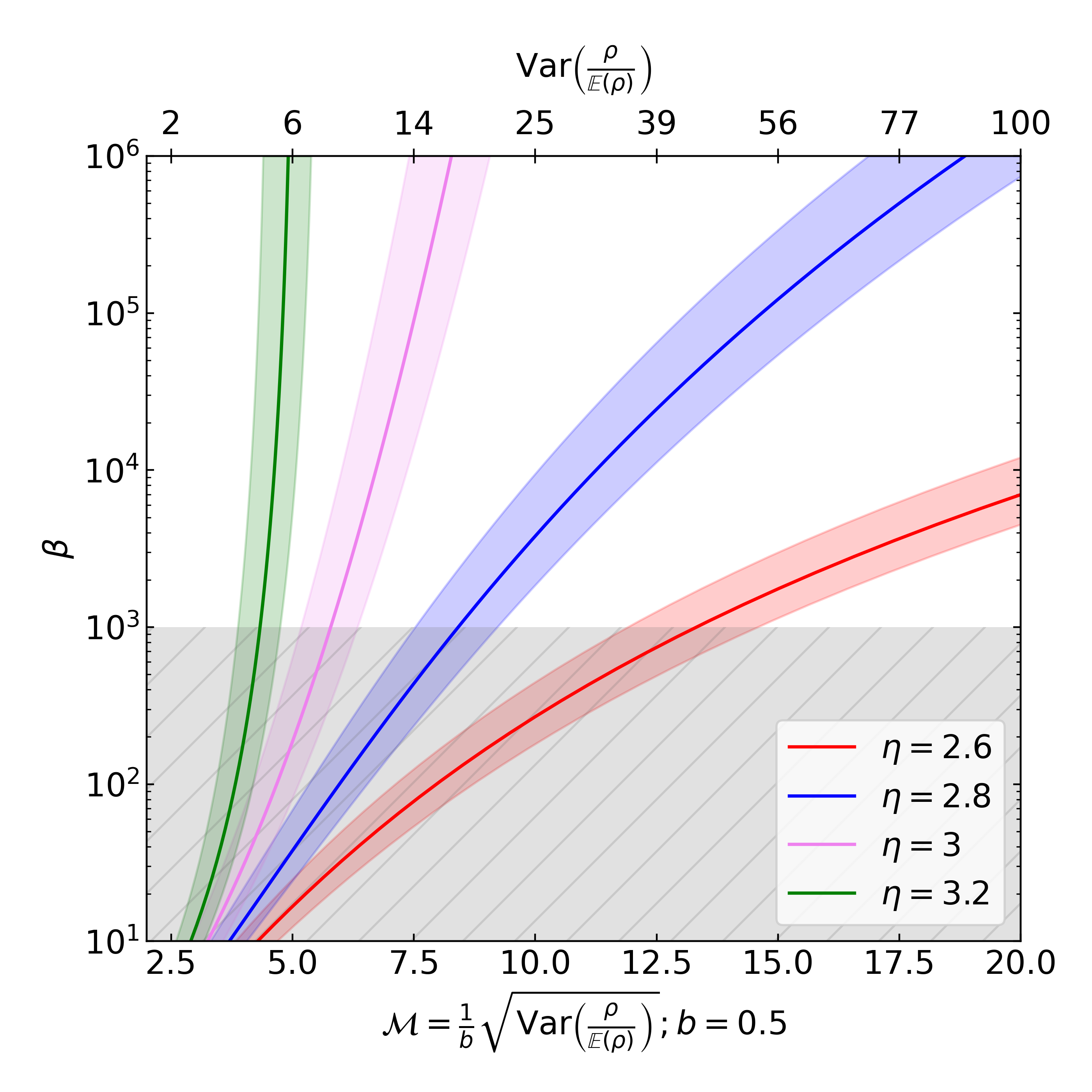}
    \caption{Inertial width, \(\beta\), against the Mach number for various slopes, \(\eta\), of the inertial range of the PS of the density field. Notably, \(\beta\) has been observed to be larger than \(10^3\) in the interstellar medium \citep{Miville-Deschenes_ProbingInterstellarTurbulence2016a, Pineda_ProbingPhysicsStarFormation2024a}, and thus we shade the excluded region in gray. The shaded areas around the solid curves represent the same as in Fig. \ref{fig_slope_model}.}
    \label{fig_slope_obs_diss}
\end{figure}

In Fig. \ref{fig_slope_obs_diss}, we plot \(\beta\) against the Mach number for various slopes of the density PS. For the above-mentioned typical star-forming conditions, we expected the slope of the PS to vary from -2.6 to -3.2. This is perfectly consistent with most observations of molecular clouds \citep{Brunt_DensityVarianceMach2010,Miville-Deschenes_HerschelSPIREObservations2010a,Hennebelle_TurbulentMolecularClouds2012, Miville-Deschenes_ProbingInterstellarTurbulence2016a,Pingel_MultiphaseTurbulenceDensity2018,Pineda_ProbingPhysicsStarFormation2024a}. Shallower slopes are also observed in clouds that are actively undergoing star formation (see \cite{Federrath_STARFORMATIONEFFICIENCY2013} for a summary). Such a behavior is also predicted from our model since the variance in such a medium is expected to be much larger than the one that would be expected in a purely turbulent medium because the self-gravity of the medium creates regions with a high density contrast.

Even when assuming a relation between the density variance and the Mach number, this study illustrates the fact that it is not possible to directly infer the Mach number from the slope of the power-law density field for \(\eta<3\) (see \citealt{Miville-Deschenes_HerschelSPIREObservations2010a,Miville-Deschenes_ProbingInterstellarTurbulence2016a,Pineda_ProbingPhysicsStarFormation2024a}) because it depends on the unknown width of the inertial range. For $\eta>3$, comparing the observed slope with the one measured in the simulations still gives a good estimate of the Mach number because the latter depends weakly on the dissipation scale. 

Our model gives a criterion on the inertial width (i.e., the resolution) that is needed to simulate a turbulent flow characterized by a given density variance and a PS with a given slope. {From Fig.~\ref{fig_slope_obs_diss}, if one wants to simulate a molecular cloud at Mach 5 with a realistic slope \(\eta=3\) \citep{Miville-Deschenes_ProbingInterstellarTurbulence2016a}, one needs a resolution allowing \(\beta\simeq 200\).}
Starting with realistic density field statistics is of prime importance because such turbulent flows are often used as initial conditions before activating gravity to study star formation in the turbulent interstellar medium \citep{Federrath_STARFORMATIONEFFICIENCY2013,Brucy_InefficientStarFormation2024a}. Starting with biased statistics of the density field will impact the statistics of the collapsed structures obtained at the end of the simulation. For example, a PS of the density field with a shallower slope will increase the density of small scale structures, making their collapse easier. Nevertheless, the precise quantification of this effect is beyond the scope of this paper.

\section{Discussion and conclusion}

We have presented a model of the slope of the PS of the density field in statistically homogeneous isotropic isothermal turbulence in numerical simulations. 
We demonstrated that this slope depends on both the variance of the density field and the width of the dissipation range. Unlike the conventional view that the inertial range is unaffected by small-scale viscous processes, our results suggest the opposite. Specifically, the density power spectrum does not exhibit an inertial range in the classical turbulent sense, in contrast to the velocity power spectrum, whose inertial range remains independent of viscosity. Mathematically, this distinction arises because the Chandrasekhar invariant constrains the density power spectrum at zero wavenumber to be constant in time, even in the presence of external forcing and viscosity. No analogous Chandrasekhar-type invariant exists for the velocity field when the flow is forced or viscous. These findings indicate that viscosity influences density structures across all scales in compressible turbulence. However, a clear physical interpretation of this mechanism remains to be established.

Thanks to this model, we were able to reproduce the measurements of the slope of the PS of the density field in turbulent simulations very well and quantitatively explain the observed trend for the flattening slope as the Mach number increases. 
The model explains the slope of the PS of the density field observed in turbulent molecular clouds. It shows that it is not possible to determine the Mach number of a molecular cloud from the observation of this slope without additional constraints on the inertial width, notably when the slope is shallower than -3. Finally, we provide a criterion for the numerical resolution that must be reached to simulate a turbulent flow characterized by a given variance and slope of the PS of the density field. 

\begin{acknowledgements} 
   We thank the anonymous referee for constructive feedback that significantly improved the quality of this work. 
   We thank Elliot Lynch and Guillaume Laibe for very helpful discussions that improved the clarity of the manuscript. PD, JF and NB were supported by the French national research agency grant ANR-23-CE31-0005 (BRIDGES). NB cheerfully thanks Edwin Santiago Leandro for his essential contribution to the 2D version of the viscosity implementation in Ramses.
\end{acknowledgements}

\bibliographystyle{aa_3authors}
\bibliography{Bib_invariant}

\begin{appendix}

\section{Model for the statistics of the log-density field}
\label{App_logdensity_PS}
In this appendix we present the statistical description of the log-density field that is used to predict the value of the invariant. Similarly to D25, the statistics of the log-density random field \(s=\ln(\rho/\mathbb{E}(\rho))\) is assumed to be characterized by a PS with the following form for low Mach numbers (\(\mathcal{M}\lesssim 5\)):
\begin{align}
    P_s^{\rm 3D}(k) = \text{Var}\left(\ln\left(\frac{\rho}{\mathbb{E}(\rho)}\right)\right) \frac{B \e^{-k \frac{L_{\rm diss}}{2\pi}}}{\left(1+\left(\frac{kL_{\rm inj}}{2\pi}\right)^2\right)^2 }.
    \label{eq_PS_logdensity_diss}
\end{align}
Here, \(B\) is a coefficient that ensures that the integral of the PS is equal to the variance of the log-density field and \(\mathbb{E}(\rho)\) is the mean density. We also introduced the dissipation scale \(L_{\rm diss}\) and the length $L_{\rm inj}$ at which most of the energy is injected in the medium\footnote{The energy is often not injected at a single wavelength, but a range of scales are excited between \(L_{\rm inj}^{\rm min}\) and \(L_{\rm inj}^{\rm max}\).}. 

This functional form has been justified numerically in \cite{Beresnyak_DensityScalingAnisotropy2005a, Brucy_InefficientStarFormation2024a} and D25. We found in our simulations that the slope of the inertial range of the PS of the log-density field is close to -3.8. The slight difference between the present parametrization (Eq. \ref{eq_PS_logdensity_diss}) and the measured slope affects the subsequent predictions of the slope of the PS of the density field by at most 5\%. A theoretical justification for an inertial range close to -4 for the log-density PS is also under investigation (Dumond et al., in prep.). 

Assuming that in the transonic limit (\(\mathcal{M}\sim 2\)), the non-gaussian features of the log-density field are limited, the statistical properties of the field are completely determined by this power spectrum. We get \(M_{\rm inv} = \alpha M_{\rm inj}\) with \(\alpha= 2 \times 10^{-2}\) and \(M_{\rm inj} = \mathbb{E}(\rho)L_{\rm inj}^3\), where \(L_{\rm inj}\) is the scale of maximum energy injection (D25).

    \section{Turbulent driving and hydrodynamical solver}
    \label{App_numerical_setup}
    
    The numerical simulations used for this study are performed with the hydrodynamical code RAMSES \citep{Teyssier_CosmologicalHydrodynamicsAdaptive2002} without adaptive mesh refinement (fixed cartersian grid). The boundary conditions are periodic. The numerical method is based on a second-order Godunov solver scheme. The solver is the Harten–Lax–van Leer-Contact (HLLC) approximate Riemann solver \citep{Toro_RiemannSolversNumerical2009}. The initial conditions consist of a fluid of atomic hydrogen at rest. The fluid follows an isothermal or polytropic equation of state. The turbulence is forced using the Ornstein-Uhlenbeck
    forcing on the acceleration \citep{Eswaran_ExaminationForcingDirect1988a,
    Schmidt_NumericalSimulationsCompressively2009}. The equations of conservation of mass and momentum that are solved are the following:
    \begin{align}
        \frac{\partial \rho}{\partial t}+\nabla \cdot(\rho \bm{{\rm v}}) & =0, \\
        \rho\left(\frac{\partial \bm{{\rm v}}}{\partial t}+(\bm{{\rm v}} \cdot \bm{\nabla}) \bm{{\rm v}}\right) & =-c_{\rm s}^2\bm{\nabla}\rho + \rho \bm{f},
    \end{align}
    where \(\rho\) is the density, \(\bm{{\rm v}}\) the velocity, \(c_{\rm s}\) the sound speed and \(f\) the turbulent driving force. 
    The viscosity is not modeled explicitly, but acts implicitly on the flow through the numerical grid viscosity. The energy equation is not solved explicitly. It is replaced by a simple polytropic equation of state \(P\propto \rho^\gamma\) with \(\gamma\) the polytropic index. In this study, we consider only the case \(\gamma=1\) (isothermal process).
    The Fourier modes $\hat{\bm{f}}(\bm{k}, t)$ of the turbulence driving acceleration field \(\bm{f}\) follow the following stochastic differential equation:
    \begin{equation}
    \mathrm{d} \hat{\bm{f}}(\bm{k}, t)=-\hat{\bm{f}}(\bm{k}, t) \frac{\mathrm{d} t}{T_{\rm OU}}+F_0(\bm{k}) \bm{P}^\zeta(\bm{k}) \mathrm{d} \bm{W}_t .
    \end{equation}
    In this equation, $\mathrm{d} t$ is the integration time step and $T_{\rm OU}$ is the autocorrelation timescale. As usually done in such numerical simulations \citep{Schmidt_NumericalSimulationsCompressively2009}, we set \(T_{\rm OU}\) to the turbulent crossing time \(T_{\rm cross}=L_{\rm inj}/\sqrt{\langle {\rm v}^2\rangle}\), where \(L_{\rm inj}\) is the injection length of turbulence and \(\sqrt{\langle {\rm v}^2\rangle}\) the turbulent velocity dispersion.
    \cite{Scannapieco_ImprovedFitDensity2025} recently showed that the variance of the density field depends on the ratio \(T_{\rm cross}/T_{\rm OU}\). However, thoughout this study, we consider only the usual case \(T_{\rm OU}=T_{\rm cross}\) letting the other cases for a future study.
    The weighting function of the driving modes $F_0$ allows the turbulence to be driven only within a precise range of spatial scales. In our work, we inject the turbulence isotropically in most runs (unless stated otherwise) between \(L_{\rm box}/6\) and \(L_{\rm box}/8\):
    \begin{equation}
        \label{eq_forcing_scale}
    F_0(k)=\left\{\begin{array}{l}
    1-\left(\frac{L_{\rm box}|\bm{k}|}{2 \pi}-7\right)^2 \text { if } 6<\frac{L_{\rm box}|\bm{k}|}{2 \pi}<8 \\
    0 \text { if not. }
    \end{array}\right.
    \end{equation}
    In most of the turbulent simulations in the literature, turbulence is injected at larger scale, typically \(L_{\rm box}/2\). Here, we chose to inject at smaller scale to ensure that the size of the simulation box is large enough compared to the correlation length \citep{Jaupart_StatisticalPropertiesCorrelation2022}. Assuming ergodicity, this ensures that spatial averages are good estimate of the statistical average.
    
    The projection operator $\bm{P}^\zeta$ is a weighted sum of the components of the Helmholtz decomposition of compressive versus solenoidal modes \citep{Federrath_ComparingStatisticsInterstellar2010}:
    \begin{align}
        P_{i j}^\zeta(k) &=\zeta P_{i j}^{\perp}(k)+(1-\zeta) P_{i j}^{\|}(k) \\
        &=\zeta \delta_{i j}+(1-2 \zeta) \frac{k_i k_j}{|k|^2}, \nonumber
    \end{align}
    where $\delta_{i j}$ is the Kronecker symbol, and $P_{i j}^{\perp}=\delta_{i j}-k_i k_j / k^2$ and $P_{i j}^{\|}=k_i k_j / k^2$ are the fully solenoidal and fully compressive projection operators, respectively. The projection operator is used to construct a purely solenoidal force field by setting the solenoidal fraction $\zeta=1$, which is used to drive turbulence in an incompressible system \citep{Eswaran_ExaminationForcingDirect1988a}. In the following we have chosen \(\zeta=0.5\), which corresponds to the energy equipartition of the velocity field: 1/3 compressive and 2/3 solenoidal.

    \section{Relation between the slope and the variance} 
    \label{App_analytical_details}
    We derive the relation given in Eq. \ref{eq_slope_variance} between the slope of the PS of the density field, \(-\eta\), the variance \(\text{Var}\left(\frac{\rho}{\mathbb{E}(\rho)}\right)\) and the inertial width \(\beta\). Starting with the functional form of the PS given in Eq. \ref{eq_fit_PS_rho}, its normalization coefficient \(A\) verifies
    \begin{equation}
        \label{eq_Parseval}
        \frac{1}{(2\pi)^3}\int_{0}^{\infty} P_\rho^{\rm 3D}(k) {\rm d}^3 k = \text{Var}(\rho),
    \end{equation}
    based on the definition of the PS that is the Fourier transform of the auto-covariance function of the density field.
    Thus,
    \begin{equation}
        \frac{1}{(2\pi)^3}\int_{0}^{\infty}  A\frac{4\pi}{\left(1+(\frac{kL_{\rm inj}^{\rm max}}{2\pi})^2\right)^\frac{\eta}{2}} \e^{-k\frac{L_{\rm diss}}{2\pi}} k^2 {\rm d} k = 1.
    \end{equation}
    By doing the variable change \(u=k L_{\rm inj}/(2\pi)\), we get
    \begin{equation}
        A = \frac{L_{\rm inj}^3}{4\pi}\left(\int_{0}^{\infty}\frac{u^2 \e^{-u\beta^{-1}}}{(1+(\phi u)^2)^\frac{\eta}{2}}{\rm d}u\right)^{-1}
    ,\end{equation}
    where \(\beta = \frac{L_{\rm inj}}{L_{\rm diss}}\) and \(\phi=L_{\rm inj}^{\rm max}/L_{\rm inj}\). In this study, \(\phi=7/6.\)
    
    The correlation length \(l_{\rm c}^\rho\) is defined as
    \begin{equation}
        l_{\rm c}^\rho = \left(\frac{1}{8C_\rho(0)}\int_{\mathbb{R}^3} C_\rho(\bm{q}) {\rm d}^3\bm{q}\right)^{1/3},
        \label{eq:lc}
    \end{equation}
    where \(C_\rho\) is the auto-covariance function of the density field. It can be related to the PS by
    \begin{align}
        \l_{\rm c}^3 &= \frac{1}{2^3\text{Var}(\rho)} P_\rho^{\rm 3D}(0), \\
        &=\frac{A}{8}, \\
        &= \frac{1}{32\pi}\frac{ L_{\rm inj}^3}{\int_{0}^{\infty}\frac{u^2 \e^{-u\beta^{-1}}}{(1+(\phi u)^2)^\frac{\eta}{2}}{\rm d}u}.
        \label{eq_lc_PS}
    \end{align}
    
    \noindent Using the value of the invariant \(M_{\rm inv}=\bar{\rho}{\rm Var}(\rho/\bar{\rho})l_{\rm c}^3\), we can relate the correlation length to the density variance. Introducing the constant \(\alpha=M_{\rm inv}/M_{\rm inj}\) where \(M_{\rm inj} = \mathbb{E}(\rho)L_{\rm inj}^3\), we have
    \begin{equation}
        \text{Var}\left(\frac{\rho}{\mathbb{E}(\rho)}\right) = \alpha\left(\frac{l_{\rm c}}{L_{\rm inj}}\right)^{-3}.
    \end{equation}Together with Eq. \ref{eq_lc_PS}, we finally find Eq. \ref{eq_slope_variance}:
    \begin{equation}
        \text{Var}\left(\frac{\rho}{\mathbb{E}(\rho)}\right) = 32\pi \alpha \int_{0}^{\infty}\frac{u^2 \e^{-u\beta^{-1}}}{(1+(\phi u)^2)^\frac{\eta}{2}}{\rm d}u.
    \end{equation}
    
    \begin{figure}
        \centering
        \includegraphics[width=\columnwidth]{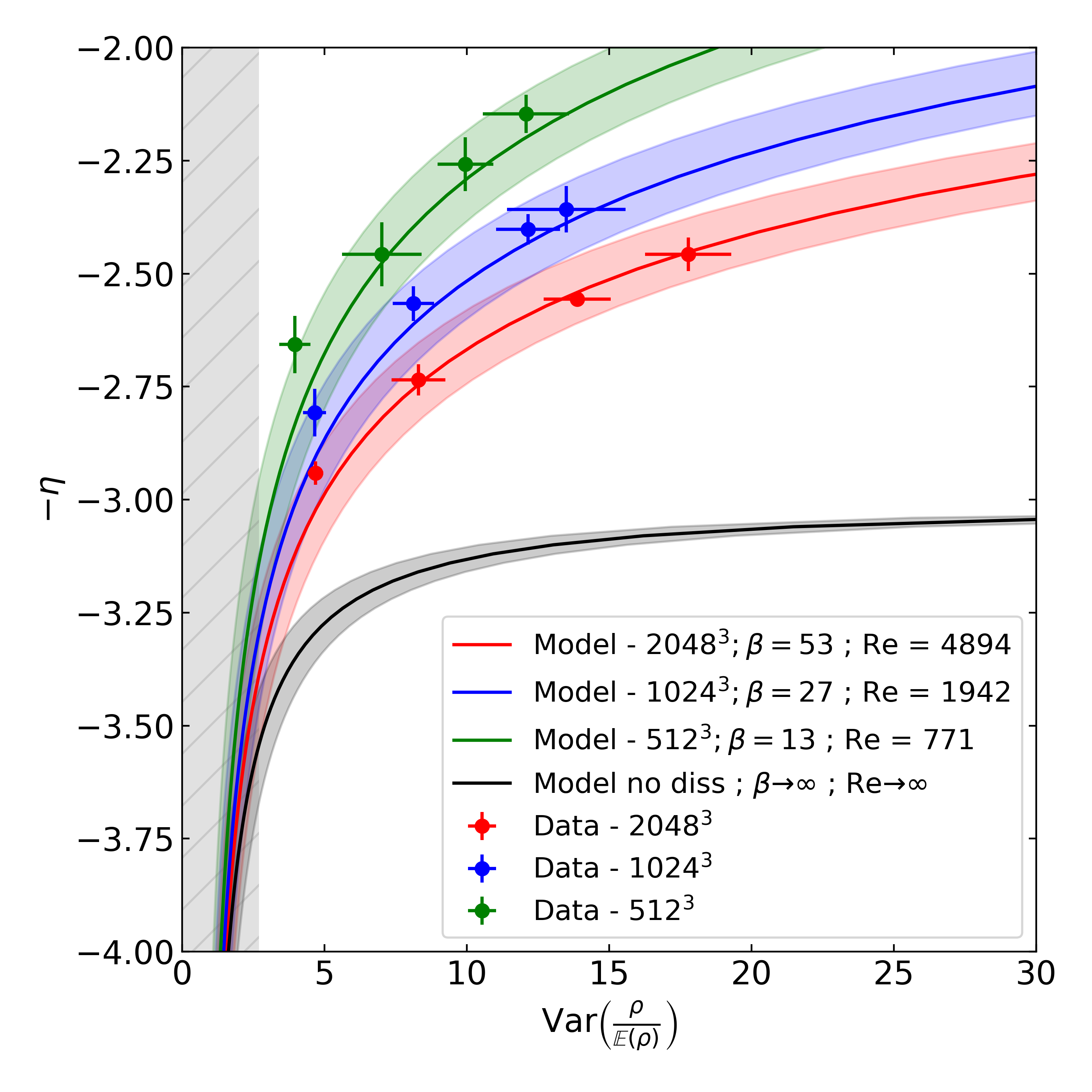}
        \caption{Same as Fig. \ref{fig_slope_model} but with the functional form of the PS of the density field given in Eq. \ref{eq_fit_PS_rho} with \(p=5\). The shaded areas around the solid curves corresponds to different predictions of the model considering an uncertainty of 20\% on the determination of the invariant around its predicted value \(\alpha=2\times 10^{-2}\). }
        \label{fig_slope_model_app}
    \end{figure}

    \section{Dependence on the functional form of the density power spectrum}
    \label{App_dependance_func_PS_rho} 
    
    In this appendix we test how the prediction of our model depends on the parameter \(p\), which control the sharpness of the transition between the injection range and the inertial range.
    
    In Fig. \ref{fig_slope_model_app} we plot the measurements of the slope in the simulations and the prediction of the model obtained with \(p=5\) in Eq.~\ref{eq_fit_PS_rho}. Because the functional form used here has changed, the values of the slopes measured in the simulation have slightly changed compared to Fig. \ref{fig_slope_model}.
    The agreement is still good, although the discrepancy between the prediction and the measurement is slightly larger at low variance. This is due to the fact that the functional form with \(p=5\) reproduces slightly less accurately the measured power spectrum at low Mach number. 
{   
   \section{Simulations with explicit viscosity}
   \label{App_explicit_visc}

   \begin{figure*}
    \centering
    \includegraphics[width=\textwidth]{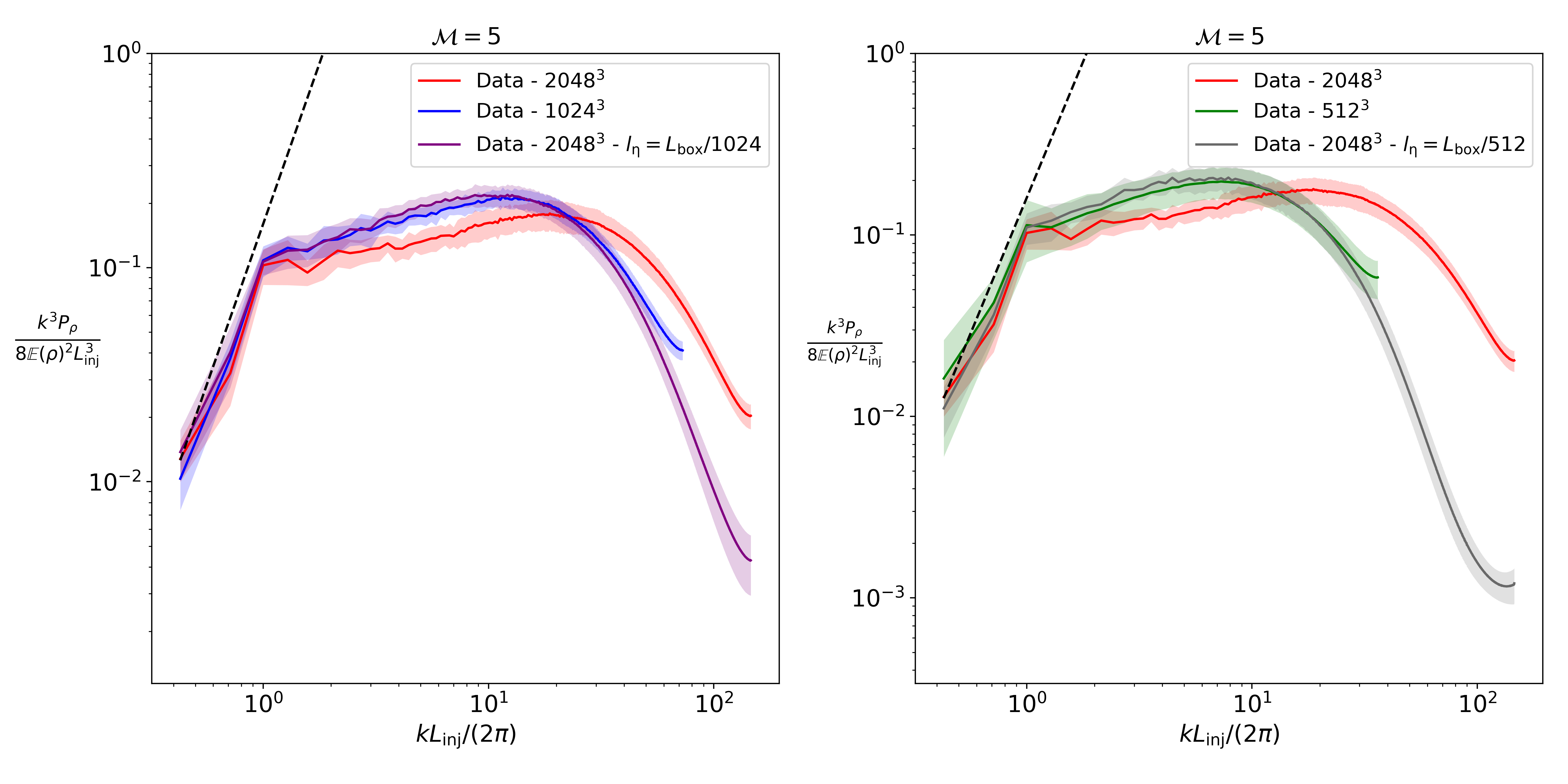}
    \caption{Comparison between the density power spectra at resolution \(1024^3\) (left) and \(512^3\) (right) with no explicit viscosity, and \(2048^3\) simulations with explicit viscosity such that the dissipation length of the compressible modes \(l_\eta\) is equal respectively to \(L_{\rm box}/1024\) and \(L_{\rm box}/512\). The power spectrum of the \(2048^3\) simulation without explicit viscosity is shown for reference. }
    \label{fig_slope_explicit_visc_app}
\end{figure*}

Since this paper suggests that the slope of the inertial range depends on the ratio between the injection scale and the dissipation scale of the density power spectrum, it is important to determine whether the dissipation scale obtained in our simulations has a physical meaning, i.e., whether it is set by the viscosity of the flow (which may be numerical if no explicit viscosity is included), or whether it is purely numerical, i.e., biased and imposed by the grid discretization. 

To address this issue, we run two additional \(2048^3\) simulations at \(\mathcal{M}=5\) with two different values of the viscosity. In order to allow comparison with the literature on viscous compressible turbulence, we use a similar model of shear viscosity to most of the studies of compressible turbulence with viscosity (e.g., \citealt{Shivakumar_NumericalViscosityResistivity2025}). The viscous stress tensor can be thus written as
\begin{equation}
    T_{\mathrm{vis}}^{i j}=\nu \rho \left(\frac{\partial v^i}{\partial x^j}+\frac{\partial v^j}{\partial x^i}-\frac{2}{3}\frac{\partial v^k}{\partial x^k} \delta^{i j}\right),
\end{equation}
where \(\nu\) is the kinematic viscosity and \(\delta^{i j}\) the Kroneker symbol. 

As shown in Fig.~\ref{fig_slope_explicit_visc_app}, increasing the explicit viscosity reduces the width of the inertial range. Moreover, it modifies the slope of the density power spectrum in a manner similar to grid viscosity, within \(1\sigma\) uncertainties. We therefore conclude that the effect of grid viscosity on the density power spectrum is comparable to that of an explicit physical viscosity.

To relate \(\beta\) to the Reynolds number, we link the dissipation length \(l_\eta\) to the size of the cells \(\Delta x\). Since the shear viscosity dissipate primarily the solenoidal modes of the velocity field, \citep{Beattie_TakingControlCompressible2025}, the dissipation length, defined as the scale at which the Reynolds number equals unity, is given by
\begin{equation}
    l_\eta = L_{\rm inj}/{\rm Re}^{3/4},
\end{equation}
where \({\rm Re} = \mathcal{M}c_{\rm s} L_{\rm inj}\nu^{-1}\). Here, \(\mathcal{M}\) is the Mach number, \(c_{\rm s}\) the sound speed, \(L_{\rm inj}\) the injection scale, and \(\nu\) the viscosity.

We verify that in a simulation with spatial resolution \(\Delta x=L_{\rm box}/N\) (with \(N^3\) the number of cells), choosing the viscosity such that the dissipation length \(l_\eta\) equals to half the cell size \(\Delta x'=L_{\rm box}/N'\) of a lower-resolution simulation (\({N'}^3\) cells, with \(N'<N\)) reproduces the dissipation scale of that simulation, as shown in Fig.~E1. We conclude that in a simulation with no explicit viscosity, the dissipation length of the compressible modes is set to half the size of the cells by the viscosity of the grid, i.e., \(l_\eta\simeq \Delta x/2\). 

We thus obtained the following relation between \(\beta\) and the Reynolds number in our numerical simulation:
\begin{equation}
    {\rm Re} = (11\beta)^{4/3} = \left(\frac{2L_{\rm inj}}{\Delta x}\right)^{4/3}
,\end{equation}
defining \(\beta=L_{\rm inj}/L_{\rm diss}\) and \(L_{\rm diss}\simeq 11l_\eta\simeq 5.5\Delta x\) in our numerical setup. This relation is used in Fig.~1 and Fig.~2 to link \(\beta\) to the Reynolds number.
}

\section{Absence of comparison to other studies}
\label{App_no_comp}

In this appendix we explain why we restrict our quantitative comparison to our own numerical simulations. Determining the slope of \(P_\rho\) requires fitting the measured density PS with the functional form given in Eq.~\ref{eq_fit_PS_rho}. An accurate fit necessitates resolving the scales larger than the injection scale, as well as a precise determination of the density variance. These conditions are not satisfied in existing simulations from the literature \citep{Kim_DensityPowerSpectrum2005,Konstandin_MachNumberStudy2016, Pan_ExactRelationDensity2022} because their turbulent forcing as done at half the box length. Moreover, when forcing at such scales, the statistical estimates of the variance and the correlation length are inaccurate (D25).
Nevertheless, our results remain qualitatively consistent with the trends reported in those studies.

\end{appendix}

\end{document}